\documentclass[12pt]{iopart}
\usepackage{epsfig}
\usepackage{iopams}  
\def\hepth#1{hep-th/#1}
\def\bfm#1{\boldsymbol{#1}}
\font\fld=msbm10 at 12 pt
\font\goth=eufm9 at 12 pt
\newcommand{\real}[1]{\mbox{\fld #1}}     
\newcommand{\gothic}[1]{\mbox{\goth #1}}    
\begin{document}
\title[Searching for new HSG theories using T-duality]{Searching for new homogeneous sine-Gordon theories using T-duality symmetries}

\author{J L Miramontes}

\address{J.~Luis Miramontes\\
Departamento de F\'\i sica de Part\'\i culas, and\\ Instituto Gallego de
F\'\i sica de Altas Energ\'\i as (IGFAE),\\Universidad
de Santiago de Compostela\\ 15782 Santiago de Compostela, Spain}
\ead{miramont@usc.es}

\begin{abstract}
The Homogeneous sine-Gordon (HSG) theories are integrable perturbations of $G_k/U(1)^{r_G}$ coset CFTs, where $G$ is a simple compact Lie group of rank $r_G$ and $k>1$ is
an integer. Using their T-duality symmetries, we investigate the relationship between the different theories corresponding to a given coset, and between the different phases of a particular theory. Our results suggest that for $G=SU(n)$ with $n\geq5$ and $E_6$ there could be two non-equivalent HSG theories associated to the same coset, one of which has not been considered so far.

\end{abstract}
\submitto{\JPA -- Special Issue~QTS5}
\maketitle

\section{Introduction}

The Homogeneous sine-Gordon (HSG) theories are two-dimensional quantum field theories with a number or remarkable properties. They are integrable perturbations of
level~$k$ $G$-parafermions, that is of coset CFTs of the form
$G_k/U(1)^{r_G}$, where $G$ is a simple compact Lie group, $k>1$ is
an integer, and $r_G$ is the rank of $G$. For simplicity, in this paper we will restrict ourselves to the cases where $G$ is simply-laced. 
The HSG theories admit a Lagrangian description in terms of a gauged Wess-Zumino-Witten (WZW) action with a potential term that was originally used to formulate them and to deduce their main features. Namely, classical integrability was proved by showing that
the classical equations of motion of the Lagrangian action can be written as zero-curvature equations in terms of an affine Kac-Moody algebra~\cite{HSG}. Quantum integrability was established using current algebra techniques after normal ordering the classical conserved densities~\cite{QHSG}. 
Moreover, the Lagrangian formulation also enabled the analysis of the semiclassical (large~$k$) limit, which provided data that were used to make conjectures for the mass spectra and exact $S$-matrices at arbitrary values of $k$~\cite{SMATHSG}. Checks of those S-matrices using both thermodynamic Bethe ansatz
(TBA) and form-factor approaches leave little doubt that they describe the perturbed parafermionic theories correctly, even for small values of $k$ far from the semiclassical regime~\cite{PATRICKTBA}. 
In particular, those checks exhibit that the theories remain quantum integrable for arbitrary (adjustable) values of $2r_G-1$ real parameters which, for particular limiting values, give rise to interesting staircase-like renormalization group (RG) trajectories. 
They also provide non-perturbative evidence for one of the most interesting features that emerge from the semiclassical studies: the presence of unstable particles.

The HSG theories are examples of two-dimensional field theories defined by actions of the
form~\cite{ZAMPCFT}
\begin{equation}
S= S_{CFT} + \mu \int d^2x\> \Phi(x)\>.
\label{Action}
\end{equation}
Here, $S_{CFT}$ denotes an action for a conformal field theory (CFT)
that governs the ultraviolet (UV) behaviour, $\mu$ is a dimensionful
coupling, and $\Phi$ is a perturbing operator. The pattern for this class of theories is provided by the perturbations of a minimal model, say ${\cal M}_p$, by its least relevant primary field, $\Phi_{1,3}$. Then, it is well known that the physical features of the resulting theory are very different for each sign of $\mu$. Namely, with $\Phi_{1,3}$ normalized as in~\cite{ZAMminimal}, for $\mu<0$ the spectrum consists of massive particles, and the theory gives rise to a RG trajectory starting in the UV fixed
point specified by ${\cal M}_p$ and flowing to a massive theory~\cite{ZAMminimal}. In contrast, for $\mu>0$ the spectrum consists of massless particles and the RG trajectory flows from the UV fixed
point specified by ${\cal M}_p$ to a non-trivial infrared (IR) fixed point 
specified by the minimal model ${\cal M}_{p-1}$~\cite{ZAMtric,MORE}. Nevertheless, this kind of behaviour is not generic, since there are many examples where the resulting theory exhibits the same features for both signs of $\mu$.

In the HSG theories, the perturbing operator has the following structure:
\begin{equation}
\mu\Phi= \sum_{i,j=1}^{r_G} \mu_i^+ \mu_j^- \> \Phi_{ij}\>,
\label{HSGpert}
\end{equation}
so that both its strength and its form depend on $2r_G-1$ adjustable parameters corresponding to $\mu_i^+$ and 
$\mu_j^-$ (notice that the perturbing operator does not change under $\mu_i^+\rightarrow \rho \mu_j^+$ and $\mu_i^-\rightarrow \rho^{-1} \mu_j^-$). In the semiclassical limit, those parameters are constrained to take values in disconnected domains labelled by the elements of ${\cal W}(G)$, the Weyl group of $G$~\cite{TDUAL}. They specify different phases of the theory where, in fact, the physical description of the spectrum of particles is different. Those domains generalize the two signs of the coupling constant in~(\ref{Action}), $\mu<0$ and $\mu>0$. Thus, taking into account the case of the perturbations of the minimal models by $\Phi_{1,3}$, this poses a question about the possibility that the features of the resulting HSG theories could be different in each phase.
A related important issue is whether each $G$ and $k$ specify a unique HSG theory. In principle, the answer might seem to be negative, because it is well known that there is not a unique CFT associated with a given coset~\cite{gWZW}. However, such possibility has not been considered in the original papers about the HSG theories.

The purpose of this paper is to address these two questions making use of target space duality. The T-duality symmetries of a family of two-dimensional massive integrable
field theories that includes the HSG theories have been described in~\cite{TDUAL}. These theories are examples of non-linear sigma
models with a potential term and, in general, T-duality provides relationships between models defined by different Lagangians which, consequently, describe the same physics.

The paper is organized as follows. In section~\ref{HSGLagrangian}, we review the main features of the HSG Lagrangian action. In particular, we discuss the form of the different actions corresponding to a given coset $G_k/U(1)^{r_G}$, which are specified by suitable automorphisms of the Cartan subalgebra of the Lie algebra of $G$. In section~\ref{Phases}, we describe the phases of the HSG theories, which are labelled by the elements of the Weyl group of $G$. The T-duality symmetries of the HSG theories are presented in section~\ref{TDualSect}, where we particularize the general construction of~\cite{TDUAL} to our case. Then, in section~\ref{SectionCentral}, we give an explicit description of the resulting symmetries for $G$ simply-laced, and we use it to show that all the phases are related by T-duality. We also investigate the possibility of having more than one non-equivalent HSG theory associated to the same coset. One of our main conclusions is that there could be up to two different HSG theories associated to $G=SU(n)$, with $n\geq5$, and $E_6$, an interesting possibility that has not been considered so far and, thus, deserves further investigation. Finally, section~\ref{Finale} contains our conclusions.

\section{The HSG Lagrangian action}
\label{HSGLagrangian}

The HSG theories corresponding to perturbations of the $G_k/U(1)^{r_G}$ coset CFT have 
actions of the form~\cite{HSG,QHSG}
\begin{equation} 
S_{\rm HSG}^{\{\tau\}}[{\gamma},A_\pm]= k\Bigl(S_{\rm gWZW}^{\{\tau\}}[{\gamma},A_\pm]
\>-\int d^2x\> V({\gamma})\Bigr)\>.
\label{ActionHSG}
\end{equation}
Here, ${\gamma}={\gamma}(t,x)$ is a
bosonic field that takes values in some faithful representation of the
compact Lie group $G$, and $A_\pm$ are non-dynamical gauge fields taking values in the Cartan subalgebra of the Lie algebra of $G$ associated with a maximal torus $H\simeq U(1)^{r_G}$ of $G$. In the following, we will denote this particular Cartan subalgebra by $\gothic{h}$, and its generators by $\bfm{h}=\{h_1,\ldots,h_{r_G}\}$.
 
In~(\ref{ActionHSG}), $kS_{\rm gWZW}^{\{\tau\}}$ is a
gauged  WZW action corresponding to the coset $G_k/U(1)^{r_G}$.
We have already anticipated that there are several CFTs associated with the same coset. Each one is specified by the group of abelian gauge transformations used in the construction of $kS_{\rm gWZW}^{\{\tau\}}$, which is of the form
\begin{equation}
\gamma\rightarrow {\rm e}^{\>i\bfm{\phi}\cdot \bfm{h}}\>\gamma\> {\rm e}^{\>-i\tau\bigl(\bfm{\phi}\bigr)\cdot \bfm{h}}\>, \qquad
A_\pm \rightarrow  A_\pm - i\partial_\pm\bfm{\phi}\cdot \bfm{h}\>,
\label{GaugeT}
\end{equation}
where $\bfm{\phi}=\bfm{\phi}(t,x)$ is a real $r_G$-dimensional vector field, and $\tau$ is a suitable automorphism of the Cartan subalgebra~\gothic{h}. Taking
$\tau=+I$ or~$-I$ leads to coset models of, so-called, vector or axial  type, respectively, while the models obtained with $\tau\not=+I,-I$ are usually called `asymmetric'~\cite{gWZW}. Consequently, the number of CFTs associated to $G_k/U(1)^{r_G}$ equals the number of possible choices of $\tau$, which has been determined in~\cite{TDUAL}.
Following this reference, an admissible automorphism $\tau$ has to satisfy two
conditions. The first one is well known: $\tau$ has to leave the restriction of the
trace form to the Cartan subalgebra~\gothic{h} invariant; namely,
$\Tr\bigl(\tau(\bfm{\phi})\cdot \bfm{h}\> \tau(\bfm{\varphi})\cdot \bfm{h}\bigr)= \Tr\bigl(\bfm{\phi}\cdot \bfm{h}\>\bfm{\varphi}\cdot \bfm{h}\bigr)$, for all $r_G$-dimensional vectors $\bfm{\phi}$ and $\bfm{\varphi}$. This is required to ensure that the group of gauge transformations~(\ref{GaugeT}) is `anomaly free'~\cite{gWZW}. Since $G$ is compact and simple, the restriction of the
trace form to \gothic{h} is (proportional to) the Euclidean
metric and, therefore, this first condition constrains
$\tau$ just to be an orthogonal $O(r_G)$ transformation~\cite{HSG,QHSG}. 

The second condition was missed in the original papers about the HSG theories. It consists in requiring that both ${\rm e}^{i\bfm{\phi}\cdot \bfm{h}}$ and ${\rm e}^{-i\tau(\bfm{\phi})\cdot \bfm{h}}$ in~(\ref{GaugeT}) belong to the same torus of $G$, which constrains $\tau$ to be an element of a discrete subgroup of  $O(r_G)$~\cite{TDUAL}. In order to spell it out, we shall make the structure of the torus $H\simeq U(1)^{r_G}$ explicit.
We start by noticing that $\exp \bigl(2\pi i \bfm{\phi}\cdot \bfm{h}\bigr)$ furnishes a map from the $r_G$-dimensional Euclidean space $\real{R}^{r_G}$, where $\bfm{\phi}$
takes values, onto $H$.  Therefore, the torus can be identified with $\real{R}^{r_G}$ factored out by the kernel of this map, which is the set of vectors $\bfm{\phi}$ mapped onto the unit element of $G$;
{\em i.e.\/}, the vectors that satisfy
\begin{equation}
\exp \Bigl(2\pi i\> \bfm{\phi}\cdot \bfm{h}\Bigr) =1\>.
\label{Monopole}
\end{equation}
This identity has to hold in any representation, and it
is convenient to write it in terms of the weights of $G$. Recall that a
weight $\bfm{\omega}\in \real{R}^{r_G}$ is the eigenvalue of
$\bfm{h}=\bigl(h_1,\ldots,h_{r_G}\bigr)$ corresponding to one common eigenvector in a single valued
representation of $G$. The set of these weights is
the `weight lattice' of $G$, which will be denoted
$\Lambda_{\rm w\/}(G)$. Then,~(\ref{Monopole}) is equivalent to $\bfm{\phi}\cdot \bfm{\omega} \in \real{Z}$ for all $\bfm{\omega}\in \Lambda_{\rm w\/}(G)$. The vectors that satisfy this condition span another lattice
$\Lambda_{\rm w\/}^\ast(G)$, known as the `dual lattice' to
$\Lambda_{\rm w\/}(G)$. Consequently, there is a solution
to~(\ref{Monopole}) for each
$\bfm{\phi}\in \Lambda_{\rm w\/}^\ast(G)$,
which provides the identification
\begin{equation}
H \simeq {\real{R}^{r_G} / \Lambda_{\rm w\/}^\ast(G)}\>.
\label{Torus}
\end{equation}
In the rather different context of gauge theories, eq.~(\ref{Monopole})
can be recognised as the quantisation condition satisfied by the magnetic
weights of monopoles, which has been solved long ago in~\cite{OLIVE}, where details about $\Lambda_{\rm
w\/}^\ast(G)$ can be found.
In particular, if $G$ is simply connected, in addition to semi-simple,
compact and connected, $\Lambda_{\rm w\/}^\ast(G)$ is the co-root lattice of
$G$, defined as the integer
span of the simple co-roots
$\bfm{\alpha}_i^\vee=(2/\bfm{\alpha}_i^2)\bfm{\alpha}_i$, where
$\{\bfm{\alpha}_1, \ldots, \bfm{\alpha}_{r_G}\}$ is a system of simple roots of $G$. It is worth noticing that, since we will only consider simply-laced Lie groups, in our case the co-root lattice coincides with the root lattice, $\Lambda_{\rm r}(G)$.
In any case, taking~(\ref{Torus}) into account, we conclude that the condition that ${\rm e}^{i\bfm{\phi}\cdot \bfm{h}}$ and ${\rm e}^{-i\tau(\bfm{\phi})\cdot \bfm{h}}$ belong to the same torus of $G$ constrains $\tau$ to be an element of the group of automorphisms of $\Lambda_{\rm w\/}^\ast(G)$; namely, 
\begin{equation} 
\tau\in{\rm Aut\;}\Lambda_{\rm w\/}^\ast(G)\>,
\label{TauCondition}
\end{equation}
which is a discrete subgroup of $O(r_G)$.

As regards the potential in~(\ref{ActionHSG}), it is given by
\begin{equation} 
V({\gamma}) =m_0^2/(4\pi) \> \Tr\left(
\Lambda_+ {\gamma}^\dagger \Lambda_- {\gamma}\right)\>,
\label{Potential}
\end{equation}
where $m_0^2$ is a bare overall mass scale and
$\Lambda_\pm=i\bfm{\lambda}_\pm\cdot\bfm{h}$ are two arbitrary constant elements
in the Cartan subalgebra~\gothic{h}.  It is straightforward to check that $V({\rm e}^{\>i\bfm{\phi}\cdot \bfm{h}}\> \gamma {\rm e}^{\>i\bfm{\varphi}\cdot \bfm{h}})= V(\gamma)$ for any $\bfm{\phi}, \bfm{\varphi}$, which exhibits that the potential is uniquely defined on the coset $G/U(1)^{r_G}$ independently of the choice of $\tau$. It is also worth noticing that~(\ref{ActionHSG}) is of the form~(\ref{Action}), with $kS_{\rm gWZW}$ and $kV$ playing the role of $S_{CFT}$ and $\mu\Phi$, respectively. Within this framework, the potential~(\ref{Potential}) is identified with a gauge invariant matrix element of the WZW field in the adjoint representation, which is a spinless relevant primary field~\cite{QHSG,KZ}.

\section{The phases of the HSG theories}
\label{Phases}

The two constant elements entering the definition of the potential, $\Lambda_+$ and $\Lambda_-$, are specified by two $r_G$-dimensional vectors, $\bfm{\lambda}_+$ and
$\bfm{\lambda}_-$. In the Lagrangian formulation, $\bfm{\lambda}_+$ and $\bfm{\lambda}_-$ play the role of $\mu_i^+$ and $\mu_i^-$ in~(\ref{HSGpert}), which shows that the potential actually depends on $2r_G-1$ adjustable real parameters (coupling constants). In order to ensure that the HSG theories have a mass gap, the choice of $\bfm{\lambda}_+$ and $\bfm{\lambda}_-$ in~(\ref{Potential}) has to be constrained such that $\bfm{\lambda}_\pm\cdot \bfm{\alpha}\not=0$ for any root $\bfm{\alpha}$ of $G$~\cite{HSG}. 

Recall that the set of hyperplanes orthogonal to all the roots of $G$
partitions the $r_G$-dimensional Euclidean space into disjoint connected components called Weyl chambers. The Weyl group of $G$ permutes the Weyl chambers, so
that each two chambers are related by a Weyl transformation. Once a system of
simple roots $\Delta=\{\bfm{\alpha}_1, \ldots,
\bfm{\alpha}_{r_G}\}$ is chosen, there is one Weyl chamber, denoted $C(\Lambda)$,
such that any $\bfm{\phi}\in C(\Lambda)$ satisfies $\bfm{\alpha}_i\cdot \bfm{\phi}\geq0$ for all
$i=1,\ldots, r_G$. $C(\Lambda)$ is called the `fundamental Weyl chamber'~\cite{HUMP}.

Consequently, since $\bfm{\lambda}_+$ and $\bfm{\lambda}_-$ cannot be orthogonal to any root of $G$, these two vectors  have to be in the interior of two (different or equal) Weyl chambers of $G$, whose choice can be used to specify the different phases of the theory. In~\cite{TDUAL}, those phases were characterized by studying the form of the manifold of vacuum field configurations, which correspond to the minima of the potential~(\ref{Potential}). The result is that there is a different phase for each element of the Weyl group of~$G$, $\sigma\in{\cal W}(G)$, which is characterized by
\begin{equation}
\bfm{\lambda}_+\in\> {\rm Int\>} C(\Delta)\quad {\rm and}\quad \bfm{\lambda}_-\in\> \sigma^{-1}\bigl(\>{\rm Int\>} C(\Delta)\bigr)\>, 
\label{PhaseCond}
\end{equation}
where ${\rm Int\>} A$ denotes the interior of the domain $A$. In the $\sigma$-phase, the vacuum confi\-gurations $\gamma_0=\gamma_\sigma$ satisfy
$\gamma_\sigma^{-1} \>\bfm{\phi}\cdot\bfm{h} \>\gamma_\sigma= \sigma(\bfm{\phi})\cdot\bfm{h}$ for all $\bfm{\phi}$.
Therefore, they are of the form
\begin{equation}
\gamma_\sigma = \widehat{\gamma}_\sigma \>{\rm e}^{\>i\bfm{\varphi}\cdot\bfm{h}}\>,
\label{VacA}
\end{equation}
where $\widehat{\gamma}_\sigma$ is fixed and $\bfm{\varphi}$ is a constant arbitrary $r_G$-dimensional vector. This shows that the space of vacuum configurations is isomorphic to $U(1)^{r_G}$. However, some of these configurations become identified
under the action of the $\tau$-dependent group of gauge
transformations~(\ref{GaugeT}). Namely,
\begin{equation}
\widehat{\gamma}_\sigma \>{\rm e}^{\>i\bfm{\varphi}\cdot\bfm{h}} \rightarrow {\rm e}^{\>i \bfm{\phi}\cdot\bfm{h}}\> \widehat{\gamma}_\sigma \>{\rm e}^{\>i\bfm{\varphi}\cdot\bfm{h}} \> {\rm e}^{\>-i \tau(\bfm{\phi})\cdot\bfm{h}} =
\widehat{\gamma}_\sigma \>{\rm e}^{\>i\bigl(\bfm{\varphi}+(\sigma-\tau)(\bfm{\phi})\bigr)\cdot\bfm{h}}\>,
\label{VacAA}
\end{equation}
which is non-trivial for each $\bfm{\phi}$ such that
$(\sigma-\tau)(\bfm{\phi})\not=0$. 
Therefore, the
manifold of physical vacuum configurations is
\begin{equation}
\{h_0 \} \simeq U(1)^{{\rm dim\; Ker\/} (\sigma-\tau)}\>,
\label{Vacb}
\end{equation}
which does depend on $\sigma$, and justifies the  identification of the phases of the HSG theories proposed in~\cite{TDUAL}. In fact, the spectrum of soliton solutions looks different in each phase. 
Notice that the action~(\ref{ActionHSG}) has a global $U(1)^{r_G}$ symmetry which, unless ${\rm dim\; Ker\/} (\sigma-\tau)=0$, does not leave the vacuum configurations invariant. Consequently, the solitons carry a $U(1)^{{\rm dim\; Im\/} (\sigma-\tau)}$ Noether charge and a $U(1)^{{\rm dim\; Ker\/} (\sigma-\tau)}$ topological charge. In particular, when there is a phase for $\sigma=+\tau$ or $\sigma=-\tau$, the corresponding solitons will be of purely topological or Noether ($Q$-ball) type, respectively, up to the charges that could be associated to the non-trivial topological properties of $G$. 

\section{Abelian T-duality in the HSG theories}
\label{TDualSect}

The HSG theories are examples of non-linear sigma models with a potential term, whose Lagrangian is of the generic form
\begin{equation} 
{\cal L}= {1\over 2}\> {\cal G}_{ij}(X)\left({\partial
X^i\over\partial t} {\partial X^j\over\partial t} - {\partial
X^i\over\partial x} {\partial X^j\over\partial x}\right) + {\cal
B}_{ij}(X)\> {\partial X^i\over\partial t} {\partial X^j\over\partial x} -
{\cal U}(X)\>,
\label{SigmaModPot}
\end{equation} 
where $i=1\ldots n$ with $n$ the dimension of the target space, ${\cal G}$ is
a metric, ${\cal B}$ is an antisymmetric tensor, and
${\cal U}$ is the potential. When ${\cal U}=0$, it is well known that each global $U(1)$ symmetry gives rise to an abelian T-duality (target space duality) symmetry  that
relates off-shell two different sigma models~\cite{Tduality}. 
A useful description of abelian T-duality in the ${\cal U}=0$ (massless) case was
originally provided by Buscher~\cite{BUSCHER}. It can be
summarized as follows. Consider the $1+1$ dimensional bosonic
non-linear sigma model defined by the Lagrangian~(\ref{SigmaModPot}) with
${\cal U}=0$. Assume that the sigma model has an abelian isometry and
that we have chosen coordinates adapted to the isometry such that it is
represented simply by a translation in the coordinate $X^1$, which
requires that ${\cal G}$ and ${\cal B}$ are independent of
$X^1$. Then, T-duality is a transformation that relates the non-linear sigma
model corresponding to $({\cal G},{\cal B})$ to another one specified by
\begin{eqnarray}
&&{\cal G}^{\rm D}_{11}={1/ {\cal G}_{11}}\>, \quad {\cal G}^{\rm D}_{1i}
={{\cal B}_{1i}/ {\cal G}_{11}}\>, \quad {\cal G}^{\rm D}_{ij}= {\cal
G}_{ij} -\bigl({{\cal G}_{1i}{\cal G}_{1j}-{\cal B}_{1i}{\cal B}_{1j}\bigr) / {\cal
G}_{11}}\>,\nonumber\\[5pt]
&&{\cal B}^{\rm D}_{1i}={{\cal G}_{1i}/ {\cal
G}_{11}}\>,
\quad {\cal B}^{\rm D}_{ij}= {\cal B}_{ij} -\bigl({{\cal G}_{1i}{\cal
B}_{1j}-{\cal B}_{1i}{\cal G}_{1j}\bigr) / {\cal G}_{11}}\>, \quad
i,j\not=1\>;
\label{Buscher}
\end{eqnarray}
moreover, $({\cal G}^{\rm D})^{\rm D}={\cal G}$ and
$({\cal B}^{\rm D})^{\rm D}={\cal B}$. Both sigma models are related by a
canonical transformation between the phase spaces that preserves the
respective Hamiltonians~\cite{CANONICAL}. Consequently, even though
they are generally defined by completely different Lagrangians, the
sigma models specified by $({\cal G},{\cal B})$ and $({\cal G}^{\rm
D},{\cal B}^{\rm D})$ turn out to describe the same physics. 

A complete description of T-duality when ${\cal U}\not=0$ is not available (for a recent discussion, see~\cite{ORL3}). However, the HSG theories belong to a particular class where T-duality symmetries arise in a particularly simple way. Namely, the class of theories whose potentials depend only on the coordinates (not on their derivatives) and preserve the abelian isometry. These potentials do not change under the canonical transformation corresponding to~(\ref{Buscher}) and, therefore, T-duality relates the models specified by
$({\cal G},{\cal B},{\cal U})$ and
$({\cal G}^{\rm D},{\cal B}^{\rm D},{\cal U})$. 
Explicit examples of integrable sigma models related by similar duality transformations are provided by the complex sine-Gordon (Lund-Regge) model~\cite{CSGDUALITY,TDUAL}, and by the models constructed in~\cite{GOMES}.

The HSG theories exhibit a global $U(1)^{r_G}$ symmetry which, in principle, provide many isometries that can be used to construct T-duality transformations. Take a particular element of the Cartan subalgebra~\gothic{h}, say $T=\bfm{t}\cdot\bfm{h}$, which generates the $U(1)$ isometry
$\gamma\rightarrow {\rm e}^{\>i\rho\> \bfm{t}\cdot\bfm{h}}\>\gamma\> {\rm e}^{\>i\rho\>\tau(\bfm{t})\cdot \bfm{h}}$,
where $\rho$ is an arbitrary real constant. According to~\cite{TDUAL}, it gives rise to a T-duality transformation of the form $({\cal G},{\cal B},{\cal U})\; \rightarrow\; ({\cal G}^{\rm D},{\cal B}^{\rm D},{\cal U})$ summarized by
\begin{equation}
S_{\rm HSG}^{\{\tau\}} \equiv S_{\rm HSG}^{\{\tau\}}[\bfm{\lambda}_+,\bfm{\lambda}_-]\;
\buildrel {\rm D_{\sigma_T}}\over{\hbox to
30pt{\rightarrowfill}} \;
S_{\rm HSG}^{\{\tau\cdot \sigma_{T}\}}[\bfm{\lambda}_+,\bfm{\lambda}_-]\>,
\label{TDual}
\end{equation}
where we have made the dependence of the HSG action on $\bfm{\lambda}_+$ and $\bfm{\lambda}_-$ explicit, and
\begin{equation}
\sigma_{T}(\bfm{\phi})= \bfm{\phi} - 2\> {\bfm{\phi}\cdot \bfm{t}\over \bfm{t}\cdot \bfm{t}}\> \bfm{t}
\label{Reflection}
\end{equation}
is just the reflection in the hyperplane orthogonal to $\bfm{t}$. Additional duality trans\-for\-mations can be constructed by performing two or more transformations one after the other, {\it i.e.\/},
\begin{equation}
S_{\rm HSG}^{\{\tau\}}[\bfm{\lambda}_+,\bfm{\lambda}_-]\;
\buildrel {\rm D_{\sigma_T}}\over{\hbox to
30pt{\rightarrowfill}}\;
S_{\rm HSG}^{\{\tau\cdot \sigma_{T}\}}[\bfm{\lambda}_+,\bfm{\lambda}_-]\;
\buildrel {\rm D_{\sigma_V}}\over{\hbox to
30pt{\rightarrowfill}}\;
S_{\rm HSG}^{\{\tau\cdot (\sigma_{T}\cdot \sigma_{V})\}}[\bfm{\lambda}_+,\bfm{\lambda}_-]
\>,
\label{TDual2}
\end{equation}
which defines $D_{\sigma_T\cdot\sigma_V}=D_{\sigma_T}\cdot D_{\sigma_V}$ for any $T,V\in\gothic{h}$. However, 
as we shall discuss in the next section, not all the resulting transformations make sense.

\section{T-duality relationships among the HSG theories}
\label{SectionCentral}

Taking~(\ref{TauCondition}) into account, the duality transformation $D_{\sigma_T}$ will only be sensible provided that $\sigma_{T}\in{\rm Aut\;}\Lambda_{\rm w\/}^\ast(G)$, which severely constrains the choice of $T$ or, equivalently, of the abelian isometry used to generate the transformation. As a consequence, the HSG theories associated with $G/U(1)^{r_G}$ will only exhibit a discrete group of abelian T-duality transformations, say ${\cal T}(G)$, spanned by the transformations corresponding to all the reflections $\sigma_{T}\in{\rm Aut\;}\Lambda_{\rm w\/}^\ast(G)$. In the next paragraphs we shall find ${\cal T}(G)$ for all the Lie groups $G$ which are simply-laced, simple, and compact.

\begin{table}[t]
\caption{\label{DynkinAut}$\Gamma_g$ for the simply-laced, simple, compact Lie groups}
\centering
\vspace{0.3 cm}
\begin{tabular}{c c c c c c c c}
\br
G & $SU(2)$ & $SU(n)$, $n\geq3$ & $SO(8)$ & $SO(2n)$, $n\geq5$ & $E_6$ & $E_7$ & $E_8$\\
\mr
$\Gamma_g$ & 1 & $\real{Z}_2$ & $S_3$ & $\real{Z}_2$ & $\real{Z}_2$ & 1 & 1 \\
\br
\end{tabular}
\end{table}

We have to distinguish two cases. The first one comprises $G=SU(n)$ with $n\geq2$, $E_6$, $E_7$ and $E_8$, which are simply connected. Then, $\Lambda_{\rm w\/}^\ast(G)$ is the (co-)root lattice of $G$ (see the comments made after~(\ref{Torus})) and, therefore, ${\rm Aut\;}\Lambda_{\rm w\/}^\ast(G)={\rm Aut\;}\Lambda_{\rm r}(G)$. 
The group of automorphisms of the root lattice is well known to be the
semidirect product of the Weyl group, ${\cal W}(G)$, and the group of
automorphisms of the Dynkin diagram, $\Gamma_g$, whose form is given in table~\ref{DynkinAut}~\cite{HUMP}. ${\cal W}(G)$ is generated by the reflections in the hyperplanes orthogonal to the roots of $G$, which means that any $\sigma\in{\cal W}(G)$ has an expression of the form
$\sigma= \sigma_{\bfm{\beta}_1}\cdots \sigma_{\bfm{\beta}_p}$, where $\bfm{\beta}_1,\ldots,\bfm{\beta}_p$ are roots of $G$ and $p\leq r_G$.
Therefore, since any $\sigma$ provides a T-duality transformation,
\begin{equation}
{\cal W}(G)\subset {\cal T}(G)\quad \forall\> G\>.
\label{WeylDual}
\end{equation}
In contrast, $\Gamma_g$ does not always give rise to sensible transformations of the form~(\ref{TDual}). Consider the case of $G=SU(n)$ with $n\geq3$. Then, the only non-trivial element of $\Gamma_g$ can be specified by its action on the basis of simple roots as follows 
\begin{equation}
w(\bfm{\alpha}_i)= \bfm{\alpha}_{n-i} \quad {\rm for}\quad i=1,\ldots,n-1\>,
\end{equation}
with the simple roots normalized such that $\bfm{\alpha}_i \cdot \bfm{\alpha}_j = 2 \delta_{ij}- \delta_{i,j-1}-\delta_{i,j+1}$. A necessary condition for this automorphism to be a reflection is that
\begin{equation}
\bfm{\alpha}_{n-1}=w(\bfm{\alpha}_1)=\sigma_T(\bfm{\alpha}_1)= \bfm{\alpha}_1 -  2\> {\bfm{\alpha}_1\cdot \bfm{t}\over \bfm{t}\cdot \bfm{t}}\> \bfm{t}\>,
\end{equation}
which singles out $\bfm{t}\propto \bfm{\alpha}_1-\bfm{\alpha}_{n-1}$. But, once $T$ is fixed,  it is straightforward to check that $\bfm{\alpha}_{n-2}= w(\bfm{\alpha}_2)=\sigma_T(\bfm{\alpha}_2) $ only for $n=3$ and $n=4$. This shows that $w$ is not a reflection for $SU(n)$ with $n\geq5$. A similar argument can be used to prove that the non-trivial element of $\Gamma_g$ is not a reflection for $E_6$ either. In other words, and from the point of view of the T-duality transformations,
\begin{eqnarray}
{\cal T}(G)={\rm Aut\;}\Lambda_{\rm w}^\ast(G)&\quad  {\rm for}\quad G=SU(n)\;\; n\leq4\>,\; E_7\>,\; E_8\nonumber\\
{\cal T}(G)={\cal W}(G)\varsubsetneq{\rm Aut\;}\Lambda_{\rm w}^\ast(G)&\quad  {\rm for}\quad G=SU(n)\;\; n\geq5\>,\; E_6\>.
\label{DynkinAutF}
\end{eqnarray}

The second case covers $G=SO(2n)$ with $n\geq3$, which is more subtle because these groups are not simply connected. We prove in the appendix that, even though $\Lambda_{\rm w}^\ast(G)\not= \Lambda_{\rm r}(G)$, for $n\not=4$ it is still true that ${\rm Aut\;}\Lambda_{\rm w}^\ast(G)$ is the
semidirect product of ${\cal W}(G)$ and $\Gamma_g$. In contrast, for $G=SO(8)$ only a $\real{Z}_2$ subgroup of $\Gamma_g$ is in ${\rm Aut\;}\Lambda_{\rm w}^\ast(G)$. 
Then, it can be easily checked that $\Lambda_{\rm w}^\ast(G)/{\cal W}(G)=\real{Z}_2$ always gives rise to sensible T-duality transformations. 
Consider the only non-trivial automorphism of this type 
which (for $n\geq3$) is given by 
\begin{eqnarray}
w(\bfm{\alpha}_i)= \bfm{\alpha}_{i} \quad {\rm for}\quad i=1,\ldots,n-2\>,\nonumber\\
w(\bfm{\alpha}_{n-1})= \bfm{\alpha}_{n}\quad {\rm and} \quad w(\bfm{\alpha}_{n})= \bfm{\alpha}_{n-1} \>,
\end{eqnarray}
with the simple roots normalized such that 
\begin{eqnarray}
\bfm{\alpha}_i\cdot \bfm{\alpha}_j&= 2 \delta_{ij}- \delta_{i,j-1}-\delta_{i,j+1}&\quad {\rm for}\quad i=1,\ldots,n-3\nonumber\\
&= 2 \delta_{ij}- \delta_{j,n-3}-\delta_{j,n-1}-\delta_{j,n} &\quad {\rm for}\quad i=n-2\nonumber\\
&= 2 \delta_{ij}- \delta_{j,n-2}&\quad {\rm for}\quad i=n-1,n\>.
\label{DnRoots}
\end{eqnarray}
Then, $w=\sigma_T$ for $\bfm{t}\propto \bfm{\alpha}_n-\bfm{\alpha}_{n-1}$ and, consequently, $w\in{\cal T}(G)$. Since the Weyl transformations always give rise to T-duality transformations, we conclude that
\begin{eqnarray}
{\cal T}(G)={\rm Aut\;}\Lambda_{\rm w}^\ast(G)&\qquad  {\rm for}\quad G=SO(n)\;\; n\geq3\>.
\label{DynkinAutF2}
\end{eqnarray}

We can now address the two questions raised in the introduction. The first one was about the relationship between the phases of a given HSG theory. Consider the phase labelled by $\sigma\in{\cal W}(G)$ which, according to~(\ref{PhaseCond}), corresponds to coupling constants of the form
\begin{equation}
\bfm{\lambda}_+\in {\rm Int\>} C(\Delta)\quad {\rm and}\quad
\bfm{\lambda}_- = \sigma^{-1}\bigl(\bfm{\lambda}_-^\bullet\bigr)\>,
\end{equation}
where $C(\Delta)$ is the principal Weyl chamber and $\bfm{\lambda}_-^\bullet\in {\rm Int\>} C(\Delta)$. Taking the form of the potential~(\ref{Potential}) into account, and making its dependence on $\bfm{\lambda}_+$ and $\bfm{\lambda}_-$ explicit, it is easy to check that $V\bigl(\gamma;\>\bfm{\lambda}_+,\bfm{\lambda}_-\bigr) = 
V\bigl(\gamma_{\sigma}^{-1}\gamma;\>\bfm{\lambda}_+,\bfm{\lambda}_-^\bullet\bigr)$,
where $\gamma_\sigma\in G$ satisfies $\gamma_\sigma^{-1} \>\bfm{\phi}\cdot\bfm{h} \>\gamma_\sigma= \sigma(\bfm{\phi})\cdot\bfm{h}$ for all $\bfm{\phi}$. Up to a change of field variables, this leads to
\begin{equation}
S_{\rm HSG}^{\{\tau\}}[\bfm{\lambda}_+,\bfm{\lambda}_-] =
S_{\rm HSG}^{\{\tau\cdot \sigma^{-1}\}}[\bfm{\lambda}_+,\bfm{\lambda}_-^\bullet]\>,
\label{Identity}
\end{equation}
which identifies the $\sigma$-phase of the HSG theory corresponding to $\tau$ with the $1$-phase of the theory corresponding to $\widetilde{\tau}=\tau\cdot\sigma^{-1}$. Then, according to~(\ref{WeylDual}), $\sigma\in{\cal W}(G)\subset {\cal T}(G)$, and we can perform the following duality transformation
\begin{equation}
S_{\rm HSG}^{\{\tau\cdot \sigma^{-1}\}}[\bfm{\lambda}_+,\bfm{\lambda}_-^\bullet]\;
\buildrel {\rm D_{\sigma}}\over{\hbox to
35pt{\rightarrowfill}}\; S_{\rm HSG}^{\{\tau\}}[\bfm{\lambda}_+,\bfm{\lambda}_-^\bullet]
\label{TDual4}
\end{equation}
which, together with~(\ref{Identity}), shows that all the phases of a given HSG theory are indeed related by the T-duality symmetries.

The second question concerned the possibility that each $G$ and $k$ could give rise to more than one non-equivalent HSG theory. The potentially different theories are constructed by considering different groups of gauge transformations of the form~(\ref{GaugeT}), which are specified by different automorphisms $\tau\in {\rm Aut\;}\Lambda_{\rm w\/}^\ast(G)$. Take two automorphisms $\tau$ and $\widetilde{\tau}$ such that $\tau^{-1}\cdot \widetilde{\tau}\in {\cal T}(G)$. Then the equations~(\ref{TDual},\ref{TDual2}) lead to
\begin{equation}
S_{\rm HSG}^{\{\tau\}}[\bfm{\lambda}_+,\bfm{\lambda}_-]\;
\buildrel D_{\tau^{-1}}\cdot D_{\widetilde{\tau}}\over{\hbox to
45pt{\rightarrowfill}}\; S_{\rm HSG}^{\{\widetilde{\tau}\}}[\bfm{\lambda}_+,\bfm{\lambda}_-]\>,
\label{TDual5}
\end{equation}
which shows that the theories corresponding to $\tau$ and $\widetilde{\tau}$ are related by T-duality. In other words, the potentially non-equivalent theories are classified by~${\rm Aut\;}\Lambda_{\rm w\/}^\ast(G)/{\cal T}(G)$.
Therefore, the answer to the question for $G$ simply-laced is provided by eqs.~(\ref{DynkinAutF}) and~(\ref{DynkinAutF2}). 
There are only two cases. The first one corresponds to $G=SU(n)$ with $n\leq4$, $SO(2n)$ with $n\geq3$, $E_7$ and $E_8$. Then, ${\cal T}(G)= {\rm Aut\;}\Lambda_{\rm w\/}^\ast(G)$ and, thus, all the resulting HSG theories are related by the T-duality symmetries discussed in section~\ref{TDualSect}. In other words, for these groups there is only one non-equivalent HSG theory associated to each coset.
The second case is more interesting. It corresponds to $G=SU(n)$ with $n\geq5$ and $E_6$. Then, ${\cal T}(G)\simeq {\cal W}(G)$ and, therefore, ${\rm Aut\;}\Lambda_{\rm w\/}^\ast(G)/{\cal T}(G)= \Gamma_g= \real{Z}_2$, which suggests that there could be up to two different HSG theories. This is one of the main results of this paper.

\section{Conclusions}
\label{Finale}

In this paper, we have used T-duality to analyze the relationships between the different phases of a particular HSG theory, and between the different HSG theories corresponding to a given coset. We have restricted our attention to simply-laced G, but we expect
that similar results will hold for the non-simply-laced cases too. The HSG theories corresponding to the coset $G/U(1)^{r_G}$ exhibit a global $U(1)^{r_G}$ symmetry that give rise to many abelian T-duality transformations. Their explicit form has been obtained by means of the particularization of the general construction of~\cite{TDUAL}. 

We have shown that T-duality relates all the phases of a given HSG theory, which means  that all of them exhibit the same physical features. We have also investigated the
relationships between the apparently different HSG theories associated to a given coset. When the group $G$ is simply-laced, in addition to simple and compact, there are only two possibilities. For $G=SU(n)$ with $n\leq4$, $SO(2n)$ with $n\geq3$, $E_7$ and $E_8$, there is only one ($k$-dependent) HSG theory, up to T-duality. In contrast, for $G=SU(n)$ with $n\geq5$ and $E_6$, the T-duality transformations suggests the possibility of constructing up to two non-equivalent theories.

The later result is rather surprising, since it has always been implicitly assumed that each $G$ and $k$ specify a unique HSG theory. Nevertheless, it is worth noticing that the usual construction of the classical and quantum HSG theories requires the existence of a phase where the vacuum
of the model is not degenerate~\cite{HSG,QHSG,SMATHSG}, which is not true for generic choices of the automorphism $\tau$ in~(\ref{GaugeT}). According to our results, the two non-equivalent theories should be constructed by considering the two groups of gauge transformations specified by the group of automorphisms of the Dynkin diagram of $G$, $\Gamma_g=\real{Z}_2$. In fact, this is equivalent to consider gauge transformations of vector ($\tau=I$) or axial ($\tau=-I$) form. Of course, there could be additional symmetries different that those considered in this paper that could relate the two resulting theories. In any case, this new possibility clearly deserves further investigation, whose starting point should be the determination of the soliton spectrum that has to be sensitive to the different structure of the vaccum field configurations of the two theories.

A final comment concerns the range of validity of the T-duality transformations discussed in section~\ref{TDualSect}. 
In principle, the Lagrangian arguments used to deduce them are expected to be valid in the semiclassical, large $k$, limit. However, 
it is worth pointing out that it has been shown, using the thermodynamic Bethe ansatz approach, that a
particular transformation of this class indeed provides a discrete symmetry of generic quantum HSG theories~\cite{TBADual}. 
This non-perturbative result supports the expectation that the semiclassical T-duality symmetries discussed in this paper provide discrete symmetries of the corresponding quantum theories, although more work is obviously required in order to validate that conjecture in general.

\vspace{0.25cm}
\ack
I would like to thank Patrick Dorey for helpful discussions, and collaborations on closely-related topics. I also thank LPTHE for hospitality.
This work is partially  supported by MEC (Spain) and FEDER (grants FPA2005-00188 and 
FPA2005-01963), by Xunta de Galicia (Conseller\'\i a de Educaci\'on and grant PGIDIT06PXIB296182PR), and by the Spanish Consolider-Ingenio 2010
Programme CPAN (CSD2007-00042).

\vspace{0.25cm}
\appendix\setcounter{section}{1}
\section*{Appendix: {\it Explicit calculation of ${\rm Aut\;}\Lambda_{\rm w\/}^\ast(G)$ for $G=SO(2n)$}}

The global structure of $G=SO(2n)$ is specified by~\cite{CORN}
\begin{equation}
SO(2n)= \widetilde{SO(2n)}/{\cal Z}\>.
\label{Global}
\end{equation}
where $\widetilde{SO(2n)}=Spin(2n)$ is the universal covering group of $SO(2n)$, which is simply connected and is uniquely determined by the Lie algebra $d_n$. Our convention for the normalization of the simple roots of $d_n$ is given in~(\ref{DnRoots}). ${\cal Z}$ is the group of order~2
\begin{equation}
{\cal Z}= \left\{1,\exp\bigl(\>\pi i (\bfm{\alpha}_{n-1}- \bfm{\alpha}_n\bigr)\cdot\bfm{h}\bigr) \right\}\simeq\real{Z}_2\>,
\end{equation}
which is a subgroup of the centre of $Spin(2n)$.

The weights of $Spin(2n)$ are easily described in terms of the set of fundamental weights $\bfm{\lambda}_1,\ldots,\bfm{\lambda}_n$ of $d_n$, which satisfy $\bfm{\alpha}_i\cdot \bfm{\lambda}_j=\delta_{ij}$. Then,
\begin{equation}
\Lambda_{\rm w\/}\bigl(Spin(2n)\bigr) =\Bigl\{\>\sum_{i=1}^n m_i\>\bfm{\lambda}_i\>;\>\; m_1,\ldots,m_n\in \real{Z}\Bigr\}\>.
\end{equation}
Recall that a weight of $G$ is the eigenvalue of
$\bfm{h}$ corresponding to one common eigenvector in a {\em single valued} representation of $G$. Consequently, since the two elements of ${\cal Z}$ become identified in $SO(2n)$,
\begin{equation}
\Lambda_{\rm w\/}(SO(2n)) =\Bigl\{\>\sum_{i=1}^n m_i\>\bfm{\lambda}_i\>;\>\; m_1,\ldots,m_n\in \real{Z},\;\;  m_{n-1}-m_n\;\; {\rm even}\Bigr\}\>.
\end{equation}
A more explicit description is obtained in terms of the orthogonal unit vectors $\bfm{e}_1,\ldots,\bfm{e}_n$, which form the usual basis of $\real{R}^n$. Then, the $2n(n-1)$ roots of $d_n$ are $\pm \bfm{e}_i \pm \bfm{e}_j$, for any $i\not=j$. The basis of simple roots reads
\begin{equation}
\bfm{\alpha}_1 = \bfm{e}_1-\bfm{e}_{2},\; \ldots, \;\;
\bfm{\alpha}_{n-1} = \bfm{e}_{n-1}-\bfm{e}_{n}, \;\;
\bfm{\alpha}_{n} = \bfm{e}_{n-1}+\bfm{e}_{n}\>,
\end{equation}
and the set of fundamental weights is
\begin{eqnarray}
\bfm{\lambda}_i= \sum_{j=1}^i \bfm{e}_j \quad {\rm for}\quad  i=1,\ldots, n-2\>,
\nonumber\\
\bfm{\lambda}_{n-1}= {1\over2}\Bigl(\>\sum_{j=1}^{n-1} \bfm{e}_j - \bfm{e}_n\> \Bigr)\>, \qquad
\bfm{\lambda}_{n}= {1\over2}\Bigl(\>\sum_{j=1}^{n-1} \bfm{e}_j + \bfm{e}_n\> \Bigr)\>.
\end{eqnarray}
Using this basis,
\begin{equation}
\Lambda_{\rm w\/}(SO(2n)) =\left\{\sum_{i=1}^n m_i\>\bfm{e}_i\>;\>\; m_1,\ldots,m_n\in \real{Z}\right\} =\Lambda_{\rm w\/}^\ast(SO(2n))\>,
\label{DualWeightDn}
\end{equation}
which is just the integer span of $\bfm{e}_1,\ldots,\bfm{e}_n$~\cite{OLIVE}. Notice that
\begin{equation}
\Lambda_{\rm r\/}(SO(2n)) =\left\{\sum_{i=1}^n m_i\>\bfm{e}_i\>;\>\; m_1,\ldots,m_n\in \real{Z},\;\;\;  \sum_{i=1}^n m_i\;\; {\rm even}\right\}\>,
\end{equation}
which is $\varsubsetneq\Lambda_{\rm w\/}^\ast(SO(2n))$.

Eq.~(\ref{DualWeightDn}) exhibits that ${\rm Aut\;}\Lambda_{\rm w\/}^\ast(SO(2n))$ is the group of permutations and sign changes of the set $\bigl\{\bfm{e}_1,\ldots,\bfm{e}_n\bigr\}$.
Recall that the Weyl group of $SO(2n)$ is the group of permutations and sign changes of this set involving only an {\em even number of signs}~\cite{HUMP}. Therefore, we conclude that ${\rm Aut\;}\Lambda_{\rm w\/}^\ast(SO(2n))$ can be described as the semidirect product of ${\cal W}(SO(2n))$ and the group generated by
\begin{equation}
\bfm{e}_n\rightarrow -\bfm{e}_n
\>, \quad {\rm and}\quad \bfm{e}_i\rightarrow \bfm{e}_i\quad {\rm for}\quad i\not=n\>,
\end{equation}
which is a subgroup of order 2 of $\Gamma_{d_n}$. Both groups coincide for $n\not=4$.

\end{document}